\documentclass[12pt,preprint]{aastex}
\usepackage{psfig,graphics}
\def\degr{\hbox{$^\circ$}}

\def\farcs{\hbox{$.\!\!^{\prime\prime}$}}         
\def\gsim{\mathrel{\hbox{\rlap{\lower.55ex \hbox {$\sim$}}
                   \kern-.3em \raise.4ex \hbox{$>$}}}}
\def\lsim{\mathrel{\hbox{\rlap{\lower.55ex \hbox {$\sim$}}
                   \kern-.3em \raise.4ex \hbox{$<$}}}}

\def\msun{\hbox{M$_{\odot}\,$}}
\def\he2{\hbox{He\,{\sc ii} $\lambda$4686}}
\def\kms{\hbox{km\,s$^{-1}\,$}}
\def\RL1{\hbox{{$R_{L_{1}}$}}}

\begin{document}  
\title{The quiescent spectrum of the AM CVn star CP Eri}
\author{P.J. Groot\altaffilmark{1}, G. Nelemans\altaffilmark{2},
D. Steeghs\altaffilmark{3}, T.R. Marsh\altaffilmark{3}}

\altaffiltext{1}{Harvard-Smithsonian Center for Astrophysics, 60
Garden Street, Cambridge, MA 02138, USA; {\tt pgroot@cfa.harvard.edu};
CfA Fellow}
\altaffiltext{2}{Astronomical Institute `Anton Pannekoek', University
of Amsterdam, Kruislaan 403, 1098 SJ Amsterdam, The Netherlands; {\tt gijsn@astro.uva.nl}}
\altaffiltext{3}{Astronomy Group, University of Southampton,
Highfield, Southampton, SO17 1BJ, UK; {\tt ds@astro.soton.ac.uk, trm@astro.soton.ac.uk}}

\begin{abstract}
We used the 6.5m MMT to obtain a spectrum of the AM CVn star CP Eri in
quiescence. The spectrum is dominated by He\,{\sc i} emission lines, which are
clearly double peaked with a peak-to-peak separation of $\sim$1900
\kms. The spectrum is similar to that of the longer period AM CVn
systems GP Com and CE 315, linking the
short and the long period AM CVn systems. In contrast with GP
Com and CE 315, the spectrum of CP Eri does not show a central 'spike'
in the line profiles, but it does show lines of Si\,{\sc ii} in
emission. The presence of these lines
indicates that the material being transferred is of higher metallicity
than in GP Com and CE 315, which, combined with the low proper motion of
the system, probably excludes a halo origin of the progenitor of CP Eri. We
constrain the primary mass to $M_1>0.27$ M$_{\odot}$ and the orbital
inclination to $33\degr<i\lsim80\degr$.
The presence of the He\,{\sc i} lines in emission opens up the
possibility for phase resolved spectroscopic studies which allows a 
determination of the system parameters and a detailed study of helium
accretion disks under highly varying circumstances. 
\end{abstract}


\setcounter{footnote}{0}
\section{Introduction}
The AM CVn stars are a heterogeneous group of nine variable stars that
are characterized by a complete lack of hydrogen and a strong
dominance of helium in their spectra (see Tab.\
\ref{tab:amcvn}). Observationally they can be roughly divided in three
groups.  First, the high state systems, AM CVn and HP Lib, that show
broad, but shallow, helium absorption lines in their spectra and 
show low-level photometric variability with periods less than 20
minutes. Second, the outburst systems (CR Boo, V803 Cen, CP Eri), 
that show large amplitude ($>$1 mag) photometric variability
on a timescale of days to weeks, as well as lower-level variability
with periods of 20-30 minutes. In their bright state these are
spectroscopically similar to the high state systems, and in their
quiescent state they show the He\,{\sc i} lines in emission, but
spectra of these systems in quiescence are rare and of low S/N. The
third category are the quiescent systems, GP Com and CE 315, that show
He\,{\sc i} emission lines and hardly any photometric variability.

It is commonly assumed that these systems are binary white dwarfs
where mass is being transferred from a very low-mass secondary
($<$0.1\msun) via a helium accretion disk to a more massive
primary. This scenario was first proposed by
Paczy\'nski (1967) and Faulkner, Flannery \& Warner (1972) to
explain the photometric flickering found in AM CVn itself by Warner \&
Robinson (1972).  Until recently, spectroscopic confirmation of this
binary scenario was only possible for GP Com, where a 46 min
spectroscopic variation was first detected by Nather, Robinson \&
Stover (1981; see also Marsh, Horne \& Rosen, 1991 and Marsh,
1999). The high state and outburst systems defied every attempt to
unveil their binary nature spectroscopically, until the detection of
an S-wave component in the He\,{\sc i} lines of AM CVn (Nelemans,
Steeghs \& Groot, 2001a). By analogy it follows that all AM CVn stars
are binaries.

The three categories can be understood as very similar binary systems
in different phases of their evolution, which proceeds from short
periods and high mass-transfer rates for the high state systems to
longer periods and lower mass-transfer rates for the quiescent systems
(e.g Warner, 1995a; Tutukov \& Yungelson, 1996; Nelemans et al.,
2001b) . This evolution is driven by the loss of angular momentum due
to gravitational wave emission.


When the mass-accretion rate is high (high state systems and
outburst systems during outburst) the accretion disks are
optically thick, leading to absorption line spectra. When the
mass-accretion rate is low (outburst systems in quiescence and the
quiescent systems) the accretion disk is optically thin, leading to
emission line spectra. A similar distinction is seen in the
hydrogen-rich Cataclysmic Variables (CVs, see e.g. Warner, 1995b).

To support the evolutionary sequence, it would be of great benefit to
show that the quiescent spectrum of the outburst systems is indeed
similar to that of GP Com and CE 315. The few quiescent spectra of the
outburst systems that are available (Abbott et al., 1992 for CP
Eri, Wood et al., 1987 for CR Boo and O'Donoghue et al., 1987 for V803
Cen) show some emission lines of He\,{\sc i} (especially He {\sc
i} $\lambda$5875), but none show a double peaked profile.

To close this gap in the spectroscopic sequence of AM CVn stars we
obtained a quiescent spectrum of the outburst system CP Eri with
the refurbished 6.5m MMT on Mt. Hopkins, AZ.

\section{CP Eri}
CP Eri was found as a faint, variable, blue star at high galactic
latitude by Luyten \& Haro (1959), who observed it at 17th magnitude,
$\sim$2.5 magnitudes brighter than its quiescent magnitude of B$\sim$19.7. A
photometric periodicity of 29 minutes was found by Howell et
al. (1991). CP Eri belongs to the outburst systems and among them
is the one with the longest orbital period, and therefore should be,
among the outbursting systems, the one that resembles GP Com and CE
315 the most.
The system was spectroscopically studied
by Abbott et al. (1992) who show the outburst spectrum to be similar
to that of the high state systems. Their quiescent spectrum shows a
blue continuum with the lines of He\,{\sc i} $\lambda$5015 and
$\lambda$5875 in emission. Although a double-peaked profile is hinted
at, the S/N ratio of the spectrum was too low to firmly establish
this. A very low S/N spectrum is also shown in Zwitter \& Munari 
(1995), but no lines are visible at all in this spectrum. 

\section{Observations}
We observed the source on the night of Dec.\ 1, 2000 with the Blue
Channel Spectrograph on the 6.5m MMT, located on Mt. Hopkins, AZ. The
500 grooves/mm grating, centered on 5200 \AA\ was used with a 1\farcs0
slit width and a 3072$\times$1024 pixel Loral CCD.  Weather conditions
were non-photometric, with scattered high clouds. Therefore no attempt
to obtain flux standards was made. The set-up resulted in an effective
spectral resolution of 3\AA\ (180 \kms at 5000\AA) over a wavelength
range of 3400-7000 \AA.  Observations were made as a sequence of three
20m exposures between 07:43-08:48 UT. HeNeAr wavelength comparison
spectra were taken at the beginning and end of each observation.

All data has been reduced using standard IRAF tasks. The spectra were
extracted using the optimal extraction routine of Horne (1986),
wavelength calibrated by using the HeNeAr comparison spectra (typical
residuals of $\sim$0.3\AA) and normalized by using a cubic spline fit
to selected wavelength regions. 

\section{The quiescent spectrum of CP Eri}
\begin{figure*}[htb]
\resizebox{14cm}{!}{\rotatebox{-90}{\includegraphics{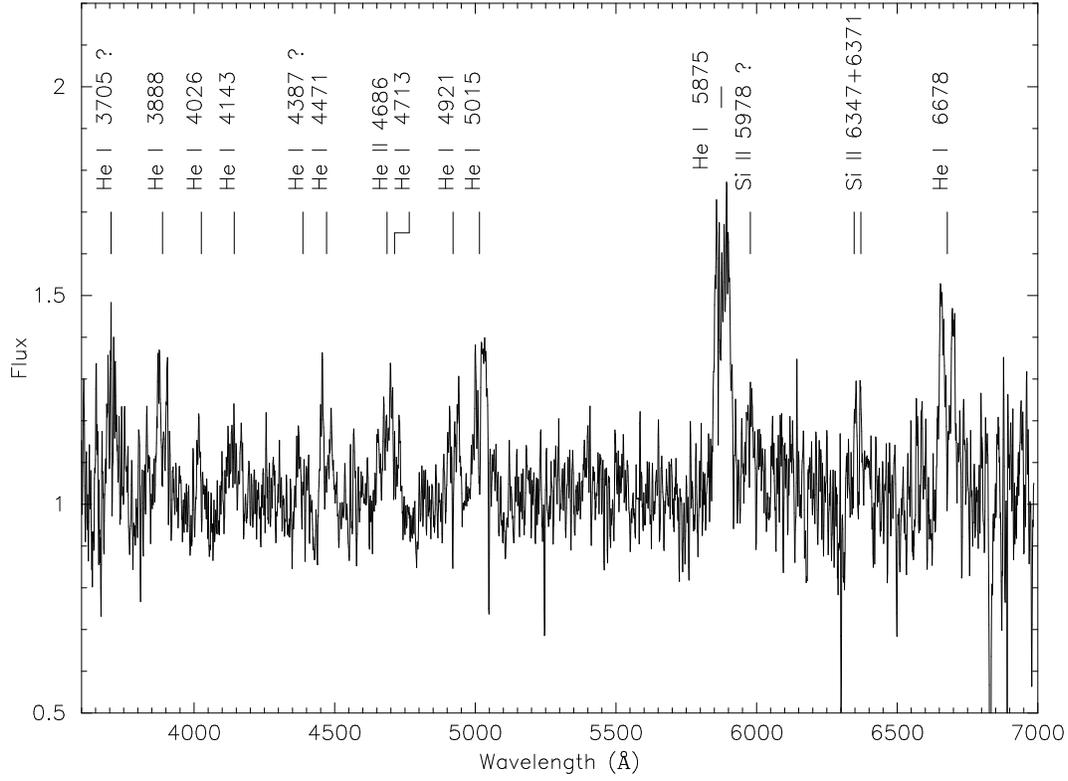}}}
\caption[]{The normalized spectrum of CP Eri in quiescence. Line
identifications are shown. \label{fig:ave}}
\end{figure*}

The median-averaged, 3-pixel box-car smoothed, quiescent spectrum of CP
Eri is shown in Fig.\ \ref{fig:ave}.  We see that it is dominated by
He\,{\sc i} emission lines, similar to GP Com and CE 315. We list the
identified lines and the equivalent widths in Table\ \ref{tab:ew}.
 
\begin{figure}[htb]
\resizebox{12cm}{!}{\rotatebox{-90}{\includegraphics{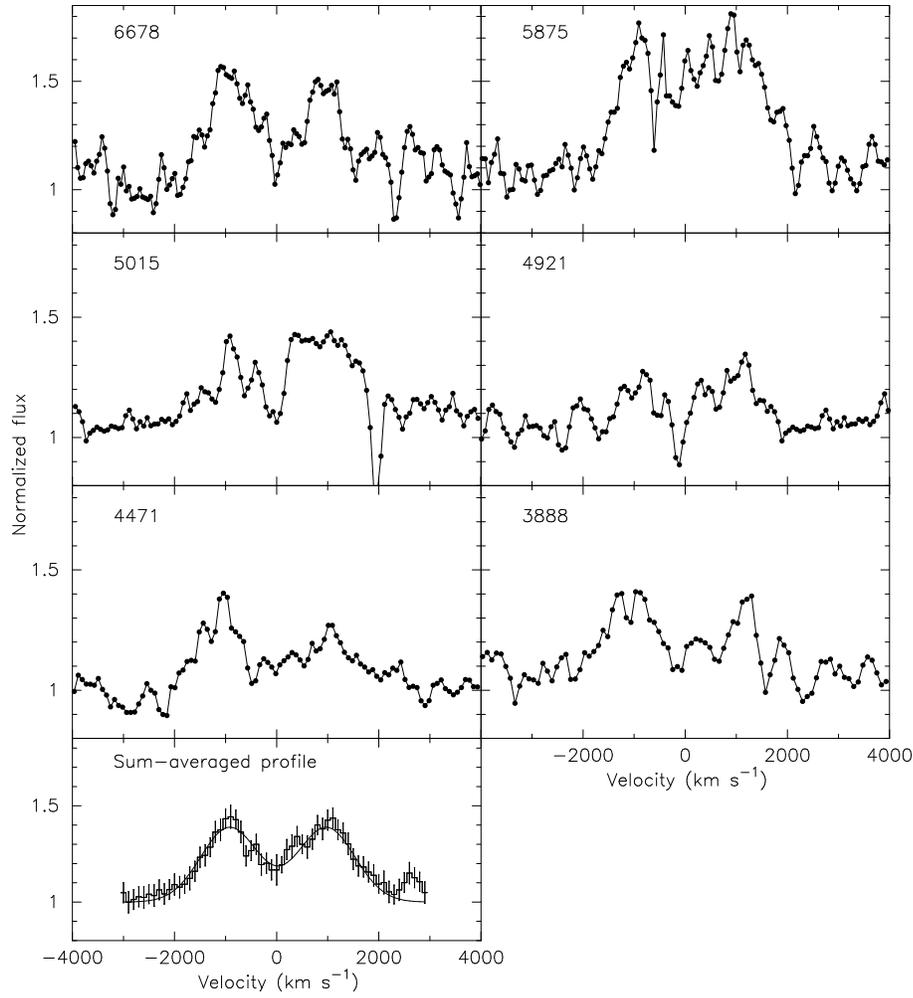}}}
\caption[]{The line profiles of the clearly identified He\,{\sc i}
lines (top panels) and the sum-averaged He\,{\sc i} line profile in 100 \kms
velocity bins (binned line, bottom panel) and a double Gaussian
profile fit to this average profile (smooth curve, bottom
panel). \label{fig:aveHeI}}
\end{figure}

All clearly identified lines are double peaked. This double peaked
profile is commonly seen, not only in GP Com and CE 315, but also in
dwarf novae and novalike CVs, and is taken as an indication of the
formation of these lines in a rotating accretion disk (see e.g. Horne
\& Marsh, 1986). The peak velocity of these profiles is a measure of
the rotational velocity of the outer parts of the accretion disk and
can therefore be used to constrain the system parameters. In order to
improve on S/N we have added (in velocity space) the profiles of the
most prominent He\,{\sc i} lines in our spectrum: He\,{\sc i}
$\lambda$6678, $\lambda$5875, $\lambda$5015, $\lambda$4921,
$\lambda$4471 and $\lambda$3888.  We did not use the He\,{\sc i}
$\lambda$4713 line because of its blend with \he2. In Fig.\
\ref{fig:aveHeI} (upper six panels) we show the line profiles of these
lines. The sum-averaged profile in 100 \kms bins (binned line) and a
double Gaussian profile fit to this sum-averaged profile is shown in
the bottom panel of Fig.\ \ref{fig:aveHeI}. For the Gaussian fit we
have used a symmetric profile where the width and height of the two
components were kept equal. The best fitted values are given in Tab.\
\ref{tab:heI}. A fit with all parameters free gave a slightly wider
red peak and similar peak velocities, but did not provide a
significantly better fit. From the asymmetry in the central velocity of
the peaks we deduce a systemic velocity, $\gamma$=23$\pm$53 \kms,
i.e. not significantly different from zero. From half of the
peak-to-peak separation of the the profile we deduce a rotational velocity of
the material in the outer disk of CP Eri of 946$\pm$52 \kms.

\section{Limits on the primary mass and inclination}
We can use the outer disk velocity of 946 \kms to set limits on the
mass of the primary star and the inclination of the system. The size
of the primary Roche lobe can be approximated by (Paczy\'{n}ski, 1967):
\begin{equation}
\RL1 = 0.462\ a\ \left(\frac{M_1}{M_1+M_2}\right)^{1/3},
\label{eq:rl1}
\end{equation}
with $M_1$ and $M_2$ the mass of the primary and secondary, and $a$
the orbital separation of the components in the binary. 

Using Kepler's third law to write $a$ in terms of the component masses
and the orbital period and collecting all numerical constants,
 we can rewrite Eq.\ \ref{eq:rl1} as :
\begin{equation}
\RL1 = 5.48 \times 10^{-5}\ P_{\rm orb}^{2/3}\ M_1^{1/3}\  {\rm m},
\end{equation}
with the orbital period, $P_{\rm orb}$, in seconds and the primary mass
in kg. 

If we assume that the gas in the outer disk is in Keplerian motion
around the primary and that the disk extends to 70\% of the primary
 Roche lobe radius, before being truncated by tidal forces,
we can equate the radius at which a Keplerian motion of 946 \kms is
reached with 70\% of the Roche lobe radius and obtain:

\begin{equation}
\frac{G\ M_1}{(v/\sin i)^2} = 3.83 \times 10^{-5}\ P^{2/3}\ M_1^{1/3},
\end{equation}
with $G, M_1, v$ and $P$ all in SI units. This can be rewritten as:

\begin{equation}
M_1 = 4.35 \times 10^8\ v^3\ P\ {\sin^{-3}i} = 0.32\ \sin^{-3}i\ {\rm M_{\odot}} 
\end{equation}

From the fact that the light curve does not show any (grazing)
eclipses (Howell, 1992), we can set an upper limit to the inclination
of $i \lsim 80\degr$, which gives a lower limit to the primary mass of 0.34
M$_{\odot}$. Since the primary mass must be lower than the
Chandrasekhar mass of 1.4 M$_{\odot}$, this sets a lower limit on the
inclination of $i>$38\degr.

If the accretion disk only reaches to 50\% of the Roche lobe radius,
as is often seen in CV dwarf novae (Harrop-Allin \& Warner, 1996) the
lower limit to the mass becomes 0.27 \msun and the lower limit to the
inclination 33\degr.

\section{Comparison with GP Com and CE 315}

The resemblance of our quiescent spectrum with that of GP Com is
remarkable, firmly establishing the connection between the long and
the short period AM CVn systems. 

Apart from the similarities with GP Com, there are also a few marked
differences. Both in GP Com (Marsh 1999) as well as in CE 315 (Ruiz et
al., 2001) a clear 'central spike' with a very low radial velocity
amplitude ($<$10 \kms) is seen in the He\,{\sc i} line profiles, which
Marsh (1999) attributes to emission from the primary white dwarf. No
such central spike is seen in the average line profile of CP Eri
(Fig.\ \ref{fig:aveHeI}).


A further difference between CP Eri and GP Com/CE 315 is the presence
in our spectrum of the Si\,{\sc ii} $\lambda$6347, 6371 lines and
possibly the Si\,{\sc ii} $\lambda$5987 line. These are not present in
the spectra of GP Com and CE 315. Marsh, Horne \& Rosen (1991) showed
that this indicates that the material in the accretion disk of GP Com
has a severely sub-solar metal abundance, indicating that the object
is probably a halo star. Marsh et al. show that for a progenitor with
solar metallicity, the strongest metal lines that should be visible
from the accretion disk are the Si\,{\sc ii} lines we see in our
spectrum of CP Eri. A preliminary comparison of the quiescent spectrum
of CP Eri with the models as used in Marsh et al. (1991) suggests that
the progenitor of the secondary currently seen in CP Eri had lower
than solar metallicity, but was certainly not as metal-poor as in GP
Com and CE 315. The current spectrum is however of too low S/N to
perform a quantative modelling and also to verify whether the Si\,{\sc
ii} lines are double-peaked and therefore originating in the
disk. However, if they do not originate in the disk they must come
from either the secondary, the primary or from circumbinary material
after being expelled from the system. In all these cases the ultimate
origin of this material is the secondary star and our conclusions on
the metallicity of the secondary's progenitor remain valid. Any
silicon 'native' to the primary will have diffused to the white dwarf
center and will not be visible on the surface.


\section{Discussion}
Understanding the evolution of AM CVn stars is of great astrophysical
importance because it touches many fields in astronomy where large
gaps in our knowledge still exist. According to evolutionary models
(e.g. Nelemans et al., 2001), to be seen today, AM CVn systems must
have survived three mass-transfer phases of which at least one was a
common-envelope phase; they could be contributors to the low-frequency
gravitational radiation background; and 'failed' AM CVn stars of the
He-family (white-dwarf plus low-mass helium star) could explode as
type Ia supernovae in an edge-lit detonation and thereby contribute up
to 25\% of the galactic SN Ia rate (see Nelemans et al., 2001b).

The detection of Si\,{\sc ii} lines in the quiescent spectrum of CP
Eri shows that its progenitor must have had an appreciably higher
metal abundance than the progenitors of GP Com and CE 315. Detecting
the metal lines also opens the possibility to constrain the
evolutionary history of these systems from the chemical composition of
the transferred material.

The proper motion of CP Eri can be derived from comparing the POSS-I
(on which the source is in outburst) and POSS-II plates: $\mu_{\rm
RA}$=6.3$\pm$0.6 mas/yr and $\mu_{\rm Dec}$=--15.0$\pm$0.6 mas/yr (Van
Kerkwijk, private communication). Together with detection of the metal
lines, which point to a higher metallicity than in GP Com and CE 315,
it seems likely that CP Eri is not a Population II object.

The detection of double-peaked emission lines in the quiescent
spectrum of CP Eri shows the physical homogeneity of the AM
CVn stars as mass-transferring white dwarf binaries, and opens
the possibility of studying the dynamics of the outbursting AM CVn
stars in greater detail. 

It will also allow for the study of helium accretion disks that
experience periodic changes from high to low mass-transfer
rates. Following the behaviour of the spectral lines during these
transitions is an important tool to track the changing physical
conditions in these unique disks, especially when these results are
compared with the hydrogen-rich disks found in many other systems,
e.g. cataclysmic variables. 
 
\vspace{1cm} {\bf Acknowledgments} We would like to thank Nelson
Caldwell for obtaining the observations, and the referee, Dr. Alon
Retter for very useful comments. PJG is supported by a
Harvard-Smithsonian CfA fellowship. GN is supported by NWO Spinoza
Grant 08-0 to E.P.J. van den Heuvel. DS is supported by a PPARC fellowship.


\references

Abbott T. M. C., Robinson E. L., Hill G. J. \& Haswell C. A., 1992, ApJ, 399, 680 

Burbidge, E.M. \& Strittmatter, P.A., 1971, ApJ 170, L39

Cropper M., Harrop-Allin M. K., Mason K. O., Mittaz J. P. D., Potter S. B. \& Ramsay G., 1998, MNRAS, 293, L57 

Faulkner J., Flannery B. P. \& Warner B., 1972, ApJ, 175, L79 

Harrop-Allin, M.K. \& Warner, B., 1996, MNRAS 279, 219

Horne K., 1986, PASP, 98, 609 

Horne K. \& Marsh T. R., 1986, MNRAS, 218, 716 

Howell S. B., 1991, IBVS, 3653 

Jha S., Garnavich P., Challis P., Kirshner R., Berlind P., 1998, IAUC,
6983 

Luyten W. J. \& Haro G., 1959, PASP, 71, 469 

Marsh T. R., 1999, MNRAS, 304, 443 

Marsh T. R., Horne K. \& Rosen S., 1991, ApJ 366, 535 

Nather R. E., Robinson E. L. \& Stover R. J., 1981, ApJ, 244, 269 

Nelemans G., Steeghs D. \& Groot P.J., 2001a, MNRAS, {\sl accepted};
astro-ph 0104220 

Nelemans G., Portegies Zwart S. F., Verbunt F. \& Yungelson L. R.,
2001b, A\&A, 368, 939

O'Donoghue D., Menzies J.W. \& Hill P.W., 1987, MNRAS, 227, 347 

O'Donoghue D. \& Kilkenny D., 1989, MNRAS, 236, 316 

O'Donoghue D., Kilkenny D., Chen A., Stobie R.S., Koen C., Warner B., Lawson, C.A, 1994, MNRAS, 271, 910 

Pacy\'nski, B., 1967, ActA, 17, 287 

Patterson, J., Sterner, E., Halpern, J.P. \& Raymond, J.C., 1992, ApJ, 384, 234 

Patterson, J., Halpern, J.P. \& Shambrook, A., 1993, ApJ 419, 803 

Ramsay, G., Cropper, M., Wu, K., Mason, K.O. \& Hakala, P., 2000, MNRAS 311, 75

Ruiz, M.T., Rojo, P.M., Garay, G. \& Maza, J.,, 2001, ApJ, {\sl accepted}, astro-ph 0103355 

Tutukov, A. V. \& Yungelson, L. R., 1996, MNRAS, 280, 1035 

Warner, B. \& Robinson, E. L., 1972, MNRAS, 159, 101 

Warner, B. 1995a, Ap\&SS 225, 249 

Warner, B., 1995b, {\sl Cataclysmic Variable Stars}, Cambridge Astroph. Ser. 28, CUP, Cambridge, UK 

Wood, M.A, {\sl et al.}, 1987, ApJ, 757, 771

Zwitter, T. \& Munari, U., 1995, A\&AS 114, 575

\begin{table*}[htb]
\caption[]{Spectroscopic charateristics of the AM CVn stars
\label{tab:amcvn}}
\begin{tabular}{llllll}
Name& $P_{\rm orb}$ (s) & $m_{\rm V}$& \multicolumn{2}{l}{Spectral
characteristics} & Ref. \\ \hline 
AM CVn& 1028.7&13.7-14.2 	&High state: &Broad, shallow He\,{\sc i}
absorption & 1-3\\		
	&	 &		& &He\,{\sc ii}
sometimes in emission & \\		
HP Lib& 1119	&13.6-13.7	& High state: &Broad, shallow He\,{\sc i} absorption & 4\\
CR Boo & 1471.3&13.0-18.0	& Outburst: & Broad,
shallow He\,{\sc i} absorption  & 5\\
	&	&		& Quiescence: & He
{\sc i} emission &5 \\
V803 Cen& 1611	&13.2-17.4 	& Outburst: & Broad, shallow
He\,{\sc i} absorption  & 6\\
	&	&		& Quiescence: & He
{\sc i} emission  & 7\\
CP Eri& 1724	&16.5-19.7	& Outburst:  & Broad,
shallow He\,{\sc i} absorption &8 \\
	&	&		& Quiescence: & Double
peaked emission He\,{\sc i}, He\,{\sc ii}& \\
	&	&		&	      & Si\,{\sc ii} emission, no 'spike' & \\
GP Com& 2790 	&15.7-16.0	& Quiescence:	&Double peaked He
{\sc i}, He\,{\sc ii} emission & 9-12\\
	&	&		& 		&N{\sc i} emission,
central spike in He\,{\sc i} &\\
CE 315& 3906 	&17.5		& Quiescence:	&Double peaked He
{\sc i}, He\,{\sc ii} emission & 13\\
	&	&		& 		&N{\sc i} emission,
central spike in He\,{\sc i} & \\
KL Dra& ?	& 16-20		& Outburst:	& Broad, shallow
He\,{\sc i} absorption & 14\\
	&	&		& Quiescence:	& ? &\\
RX J1914+24 & 569& $m_{\rm I}$=18.5	& Magnetic	& ? &\\ \hline
\end{tabular}

{\sc References.} For periods and magnitudes see Warner 1995a and
references therein and Cropper et al (1998) and Ramsay et al. (2000) 
for RXJ1914+24.
1. Patterson et al., 1992; 2. Patterson, Halpern
\& Shambrook, 1993; 3. Nelemans et al., 2001a; 4. O'Donoghue et al.,
1994; 5. Wood et al., 1987; 6. O'Donoghue \& Kilkenny, 1989;
7. O'Donoghue, Menzies \& Hill, 1987; 8. Abbott et al., 1992; 
9. Burbidge \& Strittmatter, 1971; 10. Nather, Robinson \& Stover,
1981; 11. Marsh, Horne \& Rosen, 1991; 12. Marsh, 1999; 13. Ruiz
et al., 2001; 14. Jha et al., 1998    
\end{table*}

\begin{table*}[htb]
\caption[]{Equivalent width of the emission lines in CP Eri
\label{tab:ew}}
\begin{tabular}{llr}
Line			 & Wavelength region & EW (\AA)\\ \hline
He\,{\sc i} $\lambda$3705&3661-3780& --9.5$\pm$1.8\\	
He\,{\sc i} $\lambda$3888&3837-3930& --9.3$\pm$1.4\\	
He\,{\sc i} $\lambda$4026&3977-4050& --1.7$\pm$1.2\\	
He\,{\sc i} $\lambda$4143&4086-4190& --5.4$\pm$1.4\\	
He\,{\sc i} $\lambda$4387&4334-4417& --1.5$\pm$1.2\\	
He\,{\sc i} $\lambda$4471&4433-4531& --5.3$\pm$1.4\\	
He $\lambda$4686/4713&4634-4764	   &--10.4$\pm$1.6\\ 
He\,{\sc i} $\lambda$4921&4878-4976& --6.0$\pm$1.3\\   
He\,{\sc i} $\lambda$5015&4966-5095&--14.1$\pm$1.5\\   
He\,{\sc i} $\lambda$5875&5789-5934&--32.4$\pm$1.8\\
Si\,{\sc ii} $\lambda$5978&5944-6011&--7.5$\pm$1.2\\  
Si\,{\sc ii} $\lambda$6347/71&6327-6400&--6.7$\pm$1.2\\
He\,{\sc i} $\lambda$6678&6617-6720 &--18.2$\pm$1.6\\   
\end{tabular}
\end{table*}

\begin{table}[htb]
\caption[]{Parameter values for a double Gaussian fit to the averaged summed
He\,{\sc i} line profile (all errors are 1-$\sigma$) \label{tab:heI}}
\begin{tabular}{lr}
\hline
Blue central velocity & --923$\pm$36 \kms \\
Red central velocity & 970$\pm$39 \kms\\
Gaussian width	& 794$\pm$52 \kms \\ 
Peak flux & 1.39$\pm$0.02\\
\end{tabular}
\end{table}

\clearpage

\end{document}